\documentclass[twocolumn,showpacs,preprintnumbers,amsmath,amssymb,prb,superscriptaddress]{revtex4}
\usepackage{graphicx}
\usepackage{dcolumn}
\usepackage{bm}
\newcommand{\bol}[1]{\boldsymbol #1}
\def\rnum#1{\expandafter{%
\romannumeral #1}}
\def\Rnum#1{\uppercase\expandafter{%
\romannumeral #1}}

\begin{document}


\title{Magnon bands of 
${\bol N}$-leg integer-spin antiferromagnetic systems in the 
weak-interchain-coupling regime}

\author{Masahiro Sato}
\affiliation{Synchrotron Radiation Research Center, Japan Atomic Energy
Agency, Sayo, Hyogo 679-5148, Japan and CREST Japan Science and
Technology Agency, Kawaguch, Saitama 332-0012, Japan}
%

\author{Masaki Oshikawa}
\affiliation{Institute for Solid State Physics, University of Tokyo
5-1-5 Kashiwanoha, Kashiwa 277-8581, Japan}%

\date{\today}

\begin{abstract}
Using the exact results of the $O(3)$ nonlinear sigma model (NLSM) and a few 
quantitative numerical data for integer-spin antiferromagnetic (AF) chains, 
we systematically estimate all magnon excitation energies 
of $N$-leg integer-spin AF ladders and tubes in the 
weak-interchain-coupling regime. 
Our method is based on a first-order perturbation theory 
for the strength of the interchain coupling. 
It can deal with any kind of interchain interactions, in principle. 
We confirm that results of the perturbation theory 
are in good agreement with those of a quantum Monte Carlo simulation 
and with our recent study based on a saddle-point approximation of 
the NLSM [Phys. Rev. B {\bf 72}, 104438 (2005)]. 
Our theory further supports the existence of a Haldane (gapped) phase 
even in a $d$-dimensional ($d\geq 2$) spatially anisotropic 
integer-spin AF model, if the exchange coupling in one direction 
is sufficiently strong compared with those in all the other directions. 
The strategy in this paper is applicable to other $N$-leg systems 
consisting of gapped chains which low-energy physics is exactly or 
quantitatively known.  
\end{abstract}

\maketitle

\section{\label{Intro}Introduction}
$N$-leg quantum spin systems, which we study in this paper, are a
natural extension of a purely one-dimensional (1D) spin chain.
The study of $N$-leg systems has continued for more than
a decade.~\cite{Sch,Sie,Gogo,Gia,LadderReview} 
Of course, it is generally 
more difficult to quantitatively solve and understand $N$-leg systems 
with a larger leg number $N$, 
although in (1+1)D systems, there are various powerful theoretical 
strategies (for example, conformal field theory,
bosonizations, integrability methods, exact diagonalization,
density-matrix renormalization group (DMRG), Monte Carlo methods, etc.). 
Actually, many theoretical studies for $N$-leg systems have focused on
two-leg systems, while three- or higher-leg systems have not been
thoroughly investigated.
However, some magnets with a larger $N$ 
indeed exist and have been investigated experimentally.~\cite{Azuma,Takano} 
Recently one can fabricate a few spin
tube materials~\cite{Ka-Ta,Nojiri,Mila,Mila2,Mila3,O-Y} as well as 
standard spin ladder ones: 
the former (latter) has a periodic (open) boundary condition along
the interchain (rung) direction.
These facts motivate us to study spin systems with an arbitrary leg
number. 
In addition, $N$-leg systems pose interesting questions in the
context of statistical physics: how do systems with a finite $N$ approach 
the corresponding higher-dimensional ones (infinite-$N$ systems), 
how do low-energy properties of $N$-leg systems depend on the value of 
$N$ (for instance, an even-odd character exists or not), is a
critical value of the strength of the interchain interaction 
finite or zero in an infinite-$N$ system, etc. 
These questions require systematic and quantitative predictions 
for $N$-leg systems. (The questions are partially 
resolved.~\cite{Rojo,OYA,Sie,Dell,Matsu,YTH,MS05,Sene,Matsu0405,Matsu_pre})

Now, thanks to the above-mentioned theoretical tools in (1+1)D systems, 
some simple 1D spin systems (Heisenberg chains, two-leg ladders, etc.) 
and field theories (sine-Gordon model, nonlinear sigma models, etc.) 
have been exactly or quantitatively solved. 
Such results are not only important for the deep and quantitative 
understanding of the solvable systems themselves, 
but also very useful as a starting point to study more
complicated or decorated systems such as large-$N$ ones. 
As well known, a low-energy effective theory for integer-spin
antiferromagnetic (AF) Heisenberg chains is 
the relativistic $O(3)$ nonlinear sigma model (NLSM) which is exactly solvable. 
Furthermore, the low-energy physical quantities of integer-spin AF 
chains have been numerically estimated.
In this paper, applying these established results of AF spin chains, 
we formulate the Rayleigh-Schr\"odinger-type perturbation theory for the
interchain (rung) coupling in $N$-leg integer-spin AF systems. 
As a result, all magnon dispersions are determined as a function 
of the leg number $N$, the strength of the rung coupling and the 
spin magnitude $S$, in the weak-rung-coupling regime. 
We verify the validity of our perturbation theory by comparing
results with those of quantum Monte Carlo (QMC) method.

Here, we note that quite recently we have studied the same
$N$-leg integer-spin AF systems with a weak rung coupling, based on a
different approach, the NLSM plus a saddle-point approximation (SPA)
method.~\cite{MS05,Sene} 
Although the SPA is not reliable for quantitative predictions, 
we believe that the results provide the systematic
understanding of $N$-leg integer-spin systems. 
The perturbation theory in this paper will also be useful in
verifying the validity of the NLSM plus SPA.

The outline of this paper is as follows. 
In Sec.~\ref{Perturbation}, we construct a perturbation theory for 
standard $N$-leg integer-spin AF ladders and tubes 
within the lowest order of the rung coupling. It 
leads to an analytical formula for all the magnon dispersions of 
the ladder and tube systems, elucidating several characteristics 
of low-energy excitation structure in the $N$-leg systems. 
The predicted bands are indeed consistent with those of QMC
method, and support our previous theory based on the NLSM plus SPA. 
In the next two sections, we apply 
perturbation theory to infinite-leg systems 
(higher-dimensional spin systems), and a system with a generalized rung coupling.  
In the final section, a brief summary of this paper is presented, and
the potential of perturbation theory is discussed. 
Appendixes~\ref{app1} and \ref{NLSMapp}, respectively, provide 
the solutions of eigenvalue problems of simple matrices, 
and the spin correlation functions calculated by the NLSM framework. 
They are used in Sec.~\ref{Perturbation}. 
In Appendix~\ref{Higher}, 
we shortly discuss higher-order perturbation terms
and the difficulty in them. 

\section{\label{Perturbation}Perturbation theory}
This section is the main content in this paper. 
The construction of perturbation theory and 
its basic results are presented.

\subsection{\label{model}Model}
Throughout the paper, we mainly consider 
the following Hamiltonian of standard $N$-leg integer-spin 
AF Heisenberg ladders and tubes:
\begin{subequations}
\label{N-leg} 
\begin{eqnarray}
\hat {\cal H}_{N\textrm{-leg}} &=& \hat {\cal H}_\parallel 
+\hat{\cal H}_{\perp},
\label{N-leg-1} 
\end{eqnarray}
\begin{eqnarray}
\hat {\cal H}_\parallel  &=& 
J\sum_{n=1}^N\sum_j\vec S_{n,j}\cdot\vec S_{n,j+1},
\,\,\,\,\,J>0, \label{N-leg-2}\\
\hat {\cal H}_{\perp} &=& J_\perp \sum_{n=1}^{N'}\sum_j
\vec S_{n,j}\cdot\vec S_{n+1,j},
\label{N-leg-3}
\end{eqnarray}
\end{subequations}
where $\vec S_{n,j}$ is integer-spin operator on site $j$ in $n$th
chain. The first term $\hat {\cal H}_\parallel$ is the Hamiltonian of
independent $N$ integer-spin AF chains. The second is the rung
coupling term, namely the exchange interaction between neighboring chains.
Ladders take $N'=N-1$, while tubes do $N'=N$ and 
$\vec S_{N+1,j}\equiv\vec S_{1,j}$.
A remarkable point in this model is that if the leg number $N$ is odd
and the rung coupling $J_\perp$ is positive (i.e., AF) in a tube, 
geometrical frustration along the rung exists.


\subsection{\label{review}Integer-spin antiferromagnetic chains 
and nonlinear sigma model}
As stated in the Introduction, we will treat the rung coupling 
$\hat {\cal H}_\perp$ as the perturbation term for 
the chain part $\hat {\cal H}_\parallel$. 
For such a treatment, the exact or quantitative knowledge
of the unperturbed part, the single integer-spin-$S$ AF Heisenberg chain 
$\hat {\cal H}_0=J\sum_j\vec S_j\cdot\vec S_{j+1}$, 
is necessary and essential. 
The chain $\hat {\cal H}_0$ has
never been solved exactly, but its low-energy effective 
theory, the $O(3)$ NLSM, is integrable and investigated well.
We use the well-known structure of the NLSM in order to analyze the
model~(\ref{N-leg}).
In this subsection, we review the structure and the relationship between
the NLSM and the chain $\hat {\cal H}_0$.~\cite{Hal,Frad,Auer,H-A,Ess}

The action of the $O(3)$ NLSM is given by 
\begin{eqnarray}
\label{actionNLSM}
A_{\textrm{NLSM}}&=&\int dxdt  \frac{1}{2g}
\Big[\frac{1}{v}\left(\partial_t \vec m\right)^2
-v\left(\partial_x\vec m\right)^2\Big],
\end{eqnarray}
where $\vec m(x,t)$ is the three-component vector field ($x$ and $t$
denote spatial and time coordinates, respectively), and the
constraint $\vec m^2=1$ is imposed. The coupling constant $g$ and the
velocity $v$ should be determined as a function of parameters in the
lattice system $\hat {\cal H}_0$.
Within the approximation used in the original mapping by 
Haldane,~\cite{Hal,Frad,Auer} $v\to 2SJa_0$ ($a_0$: lattice constant) 
and $g\to 2/S$, but true values of $g$ and $v$ somewhat 
deviate from these ones due to irrelevant-term effects neglected 
in the Haldane mapping (see the discussion below and Table~\ref{tab1}). 
In the NLSM picture, the original spin operator is written as 
\begin{subequations}
\label{spin}
\begin{eqnarray}
\vec S_j &\approx & (-1)^{x/a_0}S \vec m(x) + \vec u(x),
\label{spin-1}\\
\vec u &=& \frac{a_0}{gv}\,\,\vec m\times \partial_t\vec m,
\label{spin-2}
\end{eqnarray}
\end{subequations}
where $x=j a_0$, and the uniform part $\vec u$ is 
the ``angular momentum'' of the NLSM, a Noether current (symbol 
$\times$ here denotes outer product). 
For the $(1+1)$D $O(3)$ NLSM, 
the vacuum does not break any symmetries in the
theory, and the Hilbert space consists of the vacuum and 
massive triplet particles. 
Let us introduce some notations.~\cite{H-A,Ess} 
We can represent the vacuum and a
one-particle state with an $O(3)$ index $\alpha$ and a momentum $p$ as
follows:
\begin{eqnarray}
\label{state}
|0\rangle, &&|p,\alpha\rangle.
\end{eqnarray}
The index $\alpha$ ($=1,2,3$) corresponds to the 
$\alpha$th component of the field $\vec m$. 
In the spin chain, these two states 
correspond to the singlet, unique ground state (disordered spin liquid) 
and an excited state with a spin-1 magnon around the momentum 
(wave number) $k=\pi/a_0$. The energy dispersion of the triplet
particle is given by 
\begin{eqnarray}
\label{disp}
\omega(p) &=& \sqrt{\Delta^2+p^2v^2},
\end{eqnarray}
where $\Delta$ is the mass gap of the particle. 
This gap $\Delta$ should be regarded as the lowest excitation gap 
(i.e., Haldane gap) in the chain $\hat {\cal H}_0$. 
The true velocity $v$ can be determined from
the spin-1 magnon dispersion around $k=\pi/a_0$. Here, let us adopt the
relativistic normalization for the states in Eq.~(\ref{state}), 
$\langle 0|0\rangle=1$ and $\langle p,\alpha |p',\beta\rangle
=\delta_{\alpha\beta} 4\pi\omega(p)\delta(p-p')/v$.
From Eq. (\ref{state}), a state with $n$ triplet particles is
expressed as 
\begin{eqnarray}
\label{Mparticles}
|\{p,\alpha\}_n\rangle &=& 
|p_1,\alpha_1; p_2,\alpha_2;\dots;p_n,\alpha_n\rangle,
\end{eqnarray}
where the total energy and momentum are 
$E_n=\sum_{j=1}^n\omega(p_j)$ and $P_n=\sum_{j=1}^np_j$, respectively.
Using the notations in Eqs.~(\ref{state}) and (\ref{Mparticles}), and the
relativistic normalization, 
we can represent the resolution of the identity as
\begin{eqnarray}
\label{resolution}
\hat 1 &=& |0\rangle\langle 0|
+\sum_{n=1}^\infty\frac{1}{n!}\sum_{\{\alpha_j\}}
\int_{-\infty}^{\infty}\prod_{j=1}^n\frac{v dp_j}{4\pi \omega(p_j)}
\nonumber\\
&&\times |p_1,\alpha_1;\dots;p_n,\alpha_n\rangle
\langle p_1,\alpha_1;\dots;p_n,\alpha_n|
\nonumber\\
&\equiv& \sum_{j=0}^\infty\hat{\cal P}_j,
\end{eqnarray}
where $\hat{\cal P}_{j\neq 0}$ is the projection operator onto 
all the states with $j$ particles, and 
$\hat{\cal P}_0=|0\rangle\langle 0|$ 
$(\hat{\cal P}_j^2=\hat{\cal P}_j)$.

The form-factor method and symmetry arguments tell us several matrix
elements of the NLSM. A part of them, which will be used in the next
subsection, is summarized below. 
\begin{subequations}
\label{elements}
\begin{eqnarray}
\langle 0| m^\alpha(x,t)|0\rangle &=& 0,
\label{element-1}\\
\langle 0| m^\alpha(x,t)|p,\beta\rangle &= &
 \delta_{\alpha\beta}\sqrt{Z}e^{ipx-i\omega(p)t},
\label{element-2}\\
\langle 0|m^\alpha(x,t)|p,\beta;p',\gamma \rangle &=& 0,
\label{element-3}\\
\langle p, \beta| m^\alpha(x,t)|p',\gamma\rangle &=& 0,
\label{element-4}\\
\langle 0|u^\alpha(x,t)|0\rangle &=& 0,
\label{element-5}\\
\langle 0|u^\alpha(x,t)|p,\beta \rangle &=& 0,
\label{element-6}\\
\langle 0|u^\alpha(x,t)|p,\beta;p',\gamma \rangle 
&=& \textrm{can be finite},\label{element-7}\\
\langle p,\beta|u^\alpha(x,t) |p',\gamma \rangle
 &=& \textrm{can be finite}.\label{element-8}
\end{eqnarray}
\end{subequations}
Equations (\ref{element-1})¡¤(\ref{element-3}), (\ref{element-4})¡¤and 
(\ref{element-6}) are consequences of the fact that one-particle states
are odd for the symmetry operation $\vec m\to-\vec m$, 
while the vacuum $|0\rangle$, two-particle states, and $\vec u$ are even.
Since the vacuum is the singlet for the angular momentum $\vec u$, 
Eq.~(\ref{element-5}) holds (it is consistent with the fact that 
the ground state has zero magnetization in the chain $\hat {\cal H}_0$). 
These symmetry arguments are also useful 
in analyzing more complicated elements among multiparticle states. 
For the detail forms of Eqs.~(\ref{element-7}) and (\ref{element-8}), 
see, for example, Refs.~\onlinecite{A-W} and \onlinecite{S-A}.
The renormalization factor $Z$ in Eq.~(\ref{element-2}) could be
determined in the theory space of the NLSM, as a function of $g$, $v$,
and the ultraviolet cut off $\Lambda$. However, since the NLSM is
considered as the effective theory of the spin chain in the present
case, $Z$ should be fixed so that the NLSM reproduces 
low-energy properties of the chain. A proper way of fixing $Z$ is given by
the comparison of long-distance behavior of the spin correlation
function calculated by the NLSM approach and that obtained numerically. 
We explain such a comparison method in 
Appendix~\ref{NLSMapp}. 

Provided that $\Delta$, $v$, and $Z$ are accurately evaluated by 
numerical methods, 
the NLSM can quantitatively cover low-energy, long-distance properties
of the integer-spin chain $\hat {\cal H}_0$. 
Fortunately, values of $\Delta$, $v$, and $Z$ have been indeed known
in some cases. We summarize them in Table~\ref{tab1}. 
It is known that $Z$ is equal to $g$,~\cite{A-W,S-A} if the NLSM is
approximated as a model of triply degenerate massive free bosons via the SPA.
Table~\ref{tab1} shows that the true value of $Z$ in the spin-1 case is
smaller than the Haldane-mapping plus SPA value $g=2/S$. 
This deviation must originate from the interaction among magnons. 
True values of $Z$ and $v$ would
be closer to the Haldane-mapping plus SPA ones, $2/S$ and $2SJa_0$, 
respectively, with increasing $S$ because the NLSM is a semiclassical 
approach for quantum spin systems.

\begin{table}
\caption{\label{tab1}Data $(\Delta,v,Z)$ in spin-1, 2, and 3 AF 
chains $\hat {\cal H}_0$, calculated by some numerical 
methods (Refs.~\onlinecite{A-W,S-A,Qin,T-K}).}
\begin{ruledtabular}
\begin{tabular}{cccc}
Spin & $\Delta$ & $v$ & $Z$ \\
\hline
1 & $0.410J$ & $2.49Ja_0$  & $1.26$   \\
2 & $0.0892J$ & $4.65Ja_0$ &          \\
3 & $0.0100J$ &            &          \\
\end{tabular}
\end{ruledtabular}
\end{table}

\subsection{\label{lowest}Lowest-order perturbation theory}
Let us develop a Rayleigh-Schr\"odinger-type perturbation theory in
the $N$-leg system (\ref{N-leg}),
using the NLSM framework explained in Sec.~\ref{review}. 
Being interested in low-energy properties of the model
(\ref{N-leg}), we concentrate on the ground state and one-magnon states 
in the full Hilbert space of the unperturbed part 
$\hat {\cal H}_\parallel$.  
Following the notation in Sec.~\ref{review}, 
we can represent the ground state and a one-magnon state as follows:
\begin{eqnarray}
\label{N-leg_state}
|\Phi\rangle &=& \prod_{n=1}^N|0\rangle_n,\\
|p,\alpha,l\rangle &=& \Bigg[\prod_{n=1(n\neq l)}^N|0\rangle_n\Bigg]
\times |p,\alpha\rangle_l,
\end{eqnarray}
where the index $n$ ($l$) in $|0\rangle_n$ ($|p,\alpha\rangle_l$) 
means the chain number, and thus $|p,\alpha,l\rangle$ stands for a 
state with one magnon in $l$th chain. Similarly, the NLSM fields 
$\vec m$ and $\vec u$ in $l$th chain are expressed 
as $\vec m_l$ and $\vec u_l$, respectively. Here, we emphasize that 
all one-magnon states with a fixed momentum $p$
are degenerate in the decoupled system $\hat {\cal H}_\parallel$,
namely, the $3N$-fold degeneracy is present in the unperturbed
one-magnon space.

One can easily find that the first-order ground-state energy correction, 
$E_{GS}^{1\rm{st}}\equiv \langle \Phi| \hat{\cal H}_\perp|\Phi\rangle$, 
is exactly zero, because 
Eqs.~(\ref{element-1}) and (\ref{element-6}) hold. 
This result $E_{GS}^{1\rm{st}}=0$ is also
satisfied in the original lattice system framework.

We next consider one-magnon states. From Eq.~(\ref{resolution}), 
the projection operator onto all
one-magnon states in $N$ decoupled chains is given by 
\begin{eqnarray}
\label{Pro-1mag}
\hat {\cal P}_{\rm{one}} &=& \sum_{n=1}^N\sum_\alpha\int \frac{vdp}{4\pi\omega(p)}
|p,\alpha,n\rangle \langle p,\alpha,n|.
\end{eqnarray}
The effective Hamiltonian of one-magnon states 
in the first-order perturbation is 
\begin{eqnarray}
\label{eff-H}
\hat {\cal H}_\perp^{1\rm{st}} = 
\hat {\cal P}_{\rm{one}}\hat {\cal H}_\perp\hat {\cal P}_{\rm{one}}
\hspace{3cm}\\
\simeq \hat {\cal P}_{\rm{one}}\Big[
\sum_n^{N'}\int\frac{dx}{a_0}J_\perp
(S^2\vec m_n\cdot\vec m_{n+1}+\vec u_n\cdot\vec u_{n+1})\Big]
\hat {\cal P}_{\rm{one}},\nonumber
\end{eqnarray}
where $\vec m_{N+1}=\vec m_1$, $\vec u_{N+1}=\vec u_1$, and 
we dropped oscillating terms [$\propto (-1)^{x/a_0}$].
Using the matrix elements in Eq.~(\ref{elements}), we can transform 
$\hat {\cal H}_\perp^{1\rm{st}}$ as
\begin{widetext}
\begin{eqnarray}
\label{eff-H-2}
\hat {\cal H}_\perp^{1\rm{st}}&=&\sum_{s,\alpha}\sum_{t,\beta}
\int\frac{vdp_s}{4\pi\omega(p_s)}\int\frac{vdp_t}{4\pi\omega(p_t)}
|p_s,\alpha,s\rangle \langle p_s,\alpha,s|
\left( \frac{S^2J_\perp}{a_0}\sum_{n=1}^{N'}
\int dx \,\,\vec m_n\cdot\vec m_{n+1}\right)
|p_t,\beta,t\rangle \langle p_t,\beta,t|
\nonumber\\
&=&\sum_\alpha\sum_{n=1}^{N'}\int\frac{vdp}{4\pi\omega(p)}
\,\,\,\frac{S^2Zv}{2\omega(p)a_0}J_\perp\Big[
|p,\alpha,n\rangle \langle p,\alpha,n+1| + 
|p,\alpha,n+1\rangle \langle p,\alpha,n|\Big],
\end{eqnarray}
\end{widetext}
where $|p,\alpha,N+1\rangle=|p,\alpha,1\rangle$ and 
we used the formula 
$\frac{1}{2\pi}\int dx e^{i(p-p')x}=\delta(p-p')$.
The product $\vec u_n\cdot\vec u_{n+1}$ does not contribute to
the first-order perturbation. One sees that
the perturbation $\hat {\cal H}_\perp$ connects one-magnon states 
in neighboring chains. This effective Hamiltonian
provides the magnon-band splitting from the $3N$-fold degenerate
dispersion $\omega(p)$. The integrand in 
$\hat {\cal H}_\perp^{1\rm{st}}$ can be reexpressed as 
the $N\times N$ tridiagonal matrix, 
\begin{eqnarray}
\label{matrix}
\Omega_\perp(p)&=& 
\left(
\begin{array}{cccccccc}
  0        &   \omega_\perp(p)       &            &           &  \omega'      \\
\omega_\perp(p)      & 0       &   \ddots   &           &          \\
           & \ddots  &   \ddots   &  \ddots   &          \\
           &         &   \ddots   &  \ddots   &  \omega_\perp(p)     \\
\omega'         &         &            & \omega_\perp(p)     &  0     \\
\end{array}
\right),\,\,\,
\end{eqnarray}
where $\omega_\perp(p)=S^2 Z J_\perp v/(2\omega(p)a_0)$, a column 
and a row each denote a chain-number index, and thus 
$\omega'=\omega_\perp$ ($= 0$) in tubes (ladders). Appendix~\ref{app1}
shows that the eigenvalues of $\Omega_\perp$ are 
\begin{eqnarray}
\label{eigen}
\lambda_r &=& 2\omega_\perp(p)\cos k_r, \,\,\,\,\,\,
\rm{where}\\
k_r&=&\left\{
\begin{array}{ccc}
\frac{r}{N+1}\pi & r=1,\dots,N & {\rm ladders},\\
\frac{2r}{N}\pi & r=0,\dots,N-1\,\,\,({\rm mod}\,\, N) & {\rm tubes}.\\
\end{array}
\right.\nonumber
\end{eqnarray}
For the tube case, $k_r$ is interpreted as wave number 
along the rung direction. From these results, we can conclude that 
the $3N$-fold degenerate magnon band $\omega(p)$ is split into 
$N$ sets of new triply degenerate ones,
\begin{eqnarray}
\label{newband}
\omega_r^{1\rm{st}}(p)&=&\omega(p)+\lambda_r(p)\nonumber\\
&=& \sqrt{\Delta^2+p^2v^2}+\frac{S^2 Zv a_0^{-1}}
{\sqrt{\Delta^2+p^2v^2}}J_\perp\cos k_r,\,\,\,\,\,\,
\end{eqnarray}
by the first-order correction. 
Of course, the remaining triple degeneracy is attributed to 
the spin-1 triplet ($\alpha=1,2,3$). The gap of each band 
$\omega_r^{1\rm{st}}(p)$ is given by 
\begin{eqnarray}
\label{newgap}
\Delta_r=\omega_r^{1\rm{st}}(0)&=&
\Delta+\frac{S^2 Zv}{\Delta a_0}J_\perp\cos k_r.
\end{eqnarray} 
Here, it is found that the new dispersion $\omega_r^{1\rm{st}}(p)$ 
is not a relativistic form, though the contributing 
perturbation term $\sum_n\vec m_n\cdot \vec m_{n+1}$ is invariant under
the ``Lorentz'' transformation (field $m^\alpha$ is a Lorentz scalar).
This contradiction is due to the fact that 
our perturbation theory does not equivalently treat 
the energy and the momentum. However, it is resolved by the interpretation 
that the dispersion $\omega_r^{1\rm{st}}(p)$ is the first-order expansion
form of the completely relativistic dispersion,
\begin{eqnarray}
\label{trueband}
\omega_r(p) =\sqrt{(\Delta^2+p^2v^2)
+\frac{2S^2Zv}{a_0}J_\perp\cos k_r+O(J_\perp^2)}.
\end{eqnarray}
This is the {\it unique} relativistic band with the expanded form,
Eq.~(\ref{newband}). Equations (\ref{newband})-(\ref{trueband}) are 
the main results of perturbation theory.

From the discussion in this subsection, one would easily see 
that perturbation theory can treat any kind of rung couplings, 
and be applied to $N$-leg systems on a lattice with an arbitrary 
geometric structure besides ladders and tubes.


\subsection{\label{detail}Features of band splitting}
\begin{figure*}
\scalebox{0.6}{\includegraphics{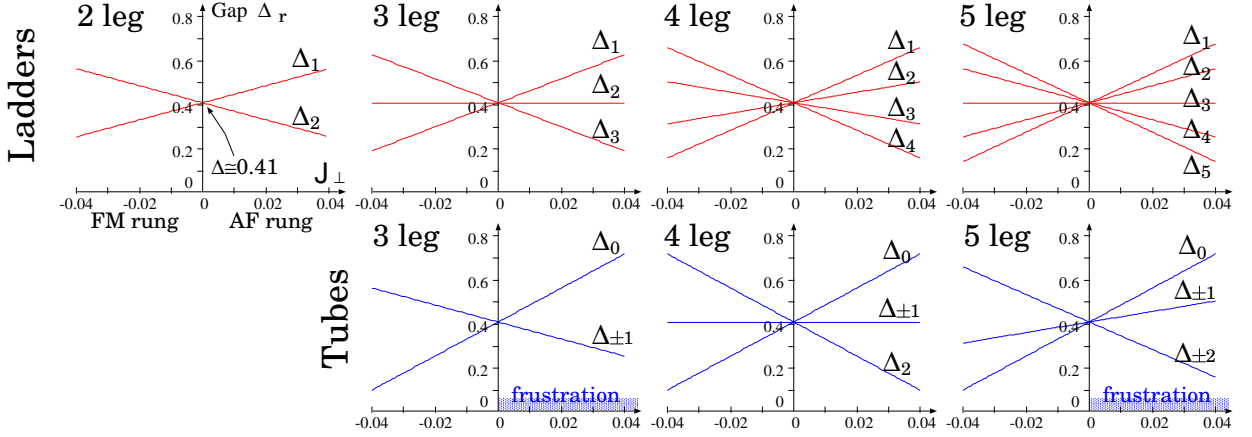}}
\caption{\label{fig_allgap}Gaps $\Delta_r$ versus rung coupling 
$J_\perp$ in the $N$-leg spin-1 AF systems~(\ref{N-leg}) with $J=1$. 
The quantity $\Delta$ is the Haldane gap of the spin-1 AF chain $\hat{\cal H}_0$.}
\end{figure*}

For the spin-1 case, that is most interesting and realistic, 
all values of $\Delta$, $v$, and $Z$ have been obtained as in
Table~\ref{tab1}. They lead to 
\begin{eqnarray}
\label{band-S1}
\omega_\perp(p)&\simeq&\frac{3.8}{\sqrt{1+37p^2a_0^2}}\,\,J_\perp.
\end{eqnarray}
Therefore, the magnon band structure for $N$-leg 
spin-1 systems can be predicted in the perturbation
theory scheme. Gaps $\Delta_r$ of spin-1 systems with several $N$ 
are summarized in Fig.~\ref{fig_allgap}. 

The new dispersion~(\ref{newband}) and Fig.~\ref{fig_allgap} 
elucidate several features of the magnon-band splitting. 

(\rnum{1}) For ladder systems ($N'=N-1$), the band splittings in 
both AF and ferromagnetic (FM) rung-coupling cases occur in
the same way. Namely, the band structure depends on the magnitude
of the rung coupling, $|J_\perp|$, but does not on the sign of 
$J_\perp$. This is because the term 
$J_\perp\sum_n\vec m_n\cdot\vec m_{n+1}$ changes its sign via the 
unitary transformation $\vec m_{n=\rm{even}}\to -\vec m_{n=\rm{even}}$, 
and the same transformation does not affect $\hat{\cal P}_{\rm{one}}$
and the effective theory for the unperturbed part, $N$ decoupled NLSMs. 
The difference between AF- and FM-rung sides 
would appear in higher-order corrections of 
the term $J_\perp\sum_n\vec u_n\cdot\vec u_{n+1}$, which is invariant
under the above transformation.   

(\rnum{2}) For the tubes ($N'=N$), in addition to the magnon triplet, 
the extra double degeneracy 
$\omega_r^{1\rm{st}}(p)=\omega_{-r}^{1\rm{st}}(p)$ is present
except for $r=0$ and $r=\frac{N}{2}$ modes 
[note that $\cos k_r=\cos(-k_r)$]. This degeneracy is attributed to the
symmetry of the $\pi$ rotation with respect to the diameter of the cross
section of tubes, i.e., the parity symmetry along the rung direction
(see Fig. 3 in Ref.~\onlinecite{MS05}). 
Therefore, it must remain even if higher-order
corrections are taken into account.

(\rnum{3}) Like the ladder case, in even-leg tubes, sign change 
$J_\perp\sum_n\vec m_n\cdot\vec m_{n+1}
\to -J_\perp\sum_n\vec m_n\cdot\vec m_{n+1}$ is possible via 
$\vec m_{n=\rm{even}}\to -\vec m_{n=\rm{even}}$. 
Therefore, the band structure in an even-leg AF-rung tube and that in
the corresponding FM-rung one are identical. 
On the other hand, the band structure in an odd-leg
AF-rung tube differs from that in the corresponding FM-rung one. 
This is because 
any transformation connecting 
AF- and FM-rung systems are absent in odd-leg tubes.

(\rnum{4}) In odd-leg tubes, the lowest excitation is 
the triply [sixfold] degenerate band $\omega_0^{1\rm{st}}(p)$ 
[$\omega_{\pm (N-1)/2}^{1\rm{st}}(p)$] for the FM-rung [AF-rung] case: 
this could refer to an even-odd character in the tube system~(\ref{N-leg}). 
The rung-coupling-driven lowest-gap reduction in odd-leg
AF-rung tubes is smaller than that in odd-leg FM-rung ones. 
The asymmetric splitting between AF- and FM-rung sides in
odd-leg tubes, and the even-odd character is due
to geometrical frustration along the rung direction in odd-leg AF-rung
tubes. 

(\rnum{5}) The lowest magnon band in FM-rung tubes ($N\geq 3$) 
$\omega_0^{1\rm{st}}(p)$ and that in even-leg AF-rung tubes 
$\omega_{N/2}^{1\rm{st}}(p)$ 
are independent of the leg number $N$. 
On the other hand (as expected), 
that in all the other systems falls down with increasing $N$. 

\begin{figure}
\scalebox{0.4}{\includegraphics{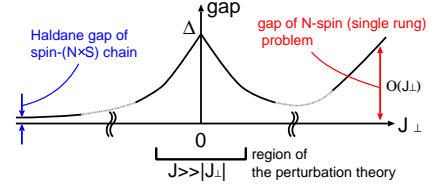}}
\caption{\label{fig_gap}Expected lowest-gap profile in the $N$-leg 
integer-spin system~(\ref{N-leg}). The symbol $\Delta$ means the 
Haldane gap of the spin-$S$ AF chain $\hat{\cal H}_0$. In the two-leg 
spin-1 case, the gap profile have already been predicted in 
Refs.~\onlinecite{Sene}, \onlinecite{All} and \onlinecite{Todo}.}
\end{figure}

(\rnum{6}) Perturbation theory shows that the lowest excitation gaps 
in all the $N$-leg systems are a monotonically decreasing 
function of $|J_\perp|$. However, as in Fig.~\ref{fig_gap}, 
from the consideration of the strong-rung-coupling regime, 
one sees that when $-J_\perp$ ($J_\perp$) is much larger than $J$, 
the lowest gap is close to the Haldane gap of the spin-$N\times S$ 
AF chain and does not vanish (the gap of the $N$ spin problem in one rung, 
which order is $J_\perp$). Predicting such a behavior would be 
beyond the scope of the present approach based on perturbation
theory in the rung coupling. 

(\rnum{7}) As we already stated in Sec.~\ref{review}, it is
expected that as $S$ is increased, $Z$ and $v$, 
respectively, become close to $g=2/S$ and $2SJa_0$. 
This result suggests the relation
\begin{eqnarray}
\label{largeS}
\lambda_r(p) &\stackrel{S\rm{: large}}{\to}&
\frac{4S^2J}{\sqrt{\Delta^2+4S^2J^2p^2a_0^2}}J_\perp\cos k_r.
\end{eqnarray}
Thus, we can infer that the band-splitting width 
monotonically increases with the growth of $S$.

The characteristic properties (\rnum{1})-(\rnum{7}) of 
the magnon-band structure in the weak-rung-coupling regime 
is completely consistent with our recent predictions based on the 
NLSM plus SPA analysis (see Sec. III A and Figs. 4-8 in 
Ref.~\onlinecite{MS05}). The property (\rnum{5}) cannot explicitly 
be derived from NLSM plus SPA, but the right panel in Fig. 7 of
Ref.~\onlinecite{MS05} implies it. 
We thus conclude that qualitative features of magnon bands predicted by 
our previous work in Ref.~\onlinecite{MS05} are supported by the present
study. For more detailed physical interpretation of the
results (\rnum{1})-(\rnum{7}), see Sec. III A of 
Ref.~\onlinecite{MS05}. The present study based on perturbation
theory furthermore provides quantitative predictions 
on band structures in the weak-rung-coupling regime. 
In particular, the analytical expressions as in
Eqs.~(\ref{newband})-(\ref{largeS}) cannot be obtained 
in the previous NLSM plus SPA approach.


\subsection{\label{compare}Quantitative comparison with 
quantum Monte Carlo Results}
Here, we compare quantitatively the lowest-excitation gaps 
of $N$-leg systems evaluated by perturbation theory and those by QMC method. 
The QMC method cannot directly calculate the excitation gaps. 
However, it may be extracted from the proper correlation functions that
can be calculated by the QMC. ~\cite{T-K,Todo,Matsu0405} 

In Table~\ref{tab2}, we summarize slopes of the lowest-excitation gaps
of $N$-leg spin-1 AF systems~(\ref{N-leg}) in the weak-rung-coupling limit, 
i.e., $-[d\Delta_{\rm min}/d|J_\perp|]_{J_\perp\to 0}$, where
$\Delta_{\rm min}$ is the minimal gap in $\{\Delta_r\}$.     
Here, the slope data of QMC method (presented by Matsumoto) 
is defined by $(\Delta_{\rm min}|_{J_\perp=0}-
\Delta_{\rm min}|_{|J_\perp|=0.01J})/(0.01J)$, where 
$\Delta_{\rm min}|_{J_\perp=0}$ is of course equal to 
the Haldane gap of the spin-1 AF Heisenberg chain $\hat{\cal H}_0$, 
$\Delta(S=1)=0.4105 J$. 
In addition to Table~\ref{tab2}, in Ref.~\onlinecite{Todo}, 
Todo and co-workers minutely investigate
the slope of the gap of the two-leg spin-1 AF ladder in the extremely
weak-rung-coupling regime, $|J_\perp|\leq 0.01J$, using QMC method 
(see Fig.~9 in Ref.~\onlinecite{Todo}). They show that the
slope is $\simeq 3.72$ in the AF-rung side, and is $\simeq 3.75$ in the
FM-rung one. This is very close to the slope predicted by 
perturbation theory, $3.8$. 
From results of Ref.~\onlinecite{Todo} and the data in Table~\ref{tab2}, 
we can conclude that perturbation theory is 
quantitatively valid in the sufficiently 
weak-rung-coupling regime, irrespective of leg number $N$.

\begin{table}
\caption{\label{tab2}Slope of lowest gaps in $N$-leg spin-1 AF
 systems~(\ref{N-leg}) near the decoupling point $J_\perp=0$, 
namely, $-[d\Delta_{\rm min}/d|J_\perp|]_{J_\perp\to0}$. 
The data of QMC is evaluated by Munehisa Matsumoto.}
\begin{ruledtabular}
\begin{tabular}{cccc}
         &        &\multicolumn{2}{c}{slope of lowest magnon gap} \\
system  &  rung  &  QMC  & perturbation theory \\
\hline
\multicolumn{1}{c}{2-leg ladder}  &\multicolumn{1}{c}{AF}  & \multicolumn{1}{c}{$3.5$} 
& \multicolumn{1}{c}{$3.8$} \\
                                  & FM &   & $3.8$   \\
\multicolumn{1}{c}{3-leg ladder}  &\multicolumn{1}{c}{AF}  & \multicolumn{1}{c}{$5.0$} 
& \multicolumn{1}{c}{$5.4$} \\
                                  & FM &   & $5.4$   \\
\multicolumn{1}{c}{3-leg tube}  &\multicolumn{1}{c}{AF}  & \multicolumn{1}{c}{} 
& \multicolumn{1}{c}{$3.8$} \\
                                  & FM &   & $7.65$   \\
\multicolumn{1}{c}{4-leg ladder}  &\multicolumn{1}{c}{AF}  & \multicolumn{1}{c}{$6.16$} 
& \multicolumn{1}{c}{$6.15$} \\
                                  & FM &   & $6.15$   \\
\multicolumn{1}{c}{4-leg tube}  &\multicolumn{1}{c}{AF}  & \multicolumn{1}{c}{$7.57$} 
& \multicolumn{1}{c}{$7.65$} \\
                                  & FM &   & $7.65$   \\
\end{tabular}
\end{ruledtabular}
\end{table}

In order to estimate the slope numerically, it is necessary to obtain
the excitation gap in high precision for small rung couplings. In 
QMC method, the precise estimate of the gap is not easy because it has
to be determined indirectly from correlation functions. On the other
hand, our perturbation theory directly gives the slope, and thus is
expected to be more reliable. Actually our prediction relies on the
numerical estimates of $\Delta$, $v$, and $Z$. However, these are
quantities defined on {\it single chains}, and can be determined
accurately by numerical calculations. Moreover, we stress that
QMC method does not work out well for {\it frustrated} (odd-leg,
AF-rung) tubes because of so-called minus-sign problem. 
The present approach has an advantage that it can be applied even to
those frustrated systems without any problem.


\section{\label{2D}${\bol N}\to \infty$ limit and Higher Dimensions}
The comparison with QMC results presented in Sec.~\ref{compare} implies 
that first-order perturbation results are efficient even for large $N$ 
(i.e., higher-order corrections are negligible at least 
in a sufficiently weak-rung-coupling case, even
if $N$ is large). Assuming this is valid even in the $N\to \infty$
limit, we can present a few predictions for
spatially anisotropic $d$-D ($d\geq 2$) spin systems, from the 
results in Sec.~\ref{Perturbation}.

In the $N\to\infty$ limit, the distribution of $k_r$ is dense.  
Therefore, the lowest-excitation modes $\omega_r^{1\rm{st}}(p)$ 
in all the infinite-leg (i.e., 2D) systems take $\cos k_r\to -1$ 
or $+1$. Particularly, for the FM-rung tube [even-leg AF-rung tube]
cases, as we already mentioned in Sec.~\ref{lowest}, the
lowest mode $\omega_0^{1\rm{st}}(p)$ [$\omega_{N/2}^{1\rm{st}}(p)$] 
always takes $\cos k_0=1$ [$\cos k_{N/2}=-1$] 
regardless of $N(\geq 3)$: namely, the lowest mode in the 2D system
already exists in a corresponding 1D one with arbitrary $N$. From 
these properties in the $N\to \infty$ case, one finds that 
the slope of the gap reduction 
$-[d\Delta_{\rm min}/d|J_\perp|]_{J_\perp\to 0}$ 
is a {\em finite} value $2\omega_\perp(0)/J_\perp$ 
in the infinite-leg system~(\ref{N-leg}).
It means that [$A$] the Haldane phase (1D gapped spin liquid) still
survives in the 2D spatially anisotropic integer-spin AF
system if the exchange coupling in the chain direction is extremely
strong, and 
[$B$] there is a finite critical value of $J_\perp$ ($J_\perp^c$) which
lies between the Haldane phase and the N\'eel ordered one. 
This prediction is consistent with results of previous 
studies.~\cite{S-T,Matsu,Tasaki,Sene2,K-K}

\begin{figure}
\scalebox{0.4}{\includegraphics{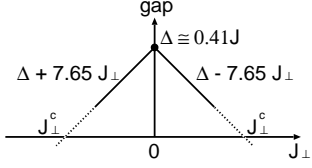}}
\caption{\label{fig_cross}Lowest excitation gap and critical point 
$J_\perp^c$ for the $N\to\infty$, spin-1 case
predicted by perturbation theory. The quantity $\Delta$ is the 
Haldane gap of the spin-1 AF chain $\hat{\cal H}_0$.}
\end{figure}

For the 2D spin-1 spatially anisotropic AF Heisenberg model 
[i.e., an infinite-leg spin-1 AF system~(\ref{N-leg}) with
$J_\perp>0$], 
the critical value $J_\perp^c$ has been numerically evaluated: 
for instance, the chain-mean-field plus exact-diagonalization 
analysis,~\cite{S-T} the cluster-expansion method,~\cite{K-K} 
and the QMC simulation~\cite{Matsu} conclude
$J_\perp^c\agt 0.025J$, $\simeq 0.056J$, and $\simeq 0.04365J$, 
respectively. On the other hand, perturbation theory provides 
the slope of the gap $2\omega_\perp(0)/J_\perp\simeq 7.65$ 
for the spin-1, $N\to\infty$ case.
As we explain in Fig.~\ref{fig_cross}, a naive extrapolation of the
linear behavior gives a vanishing of the gap at a finite $J_\perp$. 
This may be identified with the critical value, within the present
framework. 
It leads to $|J_\perp^c|\simeq 0.05J$. 
This is again consistent with the
above numerical estimates. We emphasize 
that the critical values $J_\perp^c$ for both AF- and FM-rung cases 
have the same magnitude in the first-order perturbation theory. 
This prediction also agrees with a recent QMC result.~\cite{Matsu_pre}

These results and those in Sec.~\ref{Perturbation} 
indicate that perturbation theory is quantitatively
reliable for any $N$-leg integer-spin AF system, 
including the $N\to\infty$ case.

As one readily expects, the above discussion of the 2D system can be
generalized to $d$-D lattice cases.
Namely, following the similar calculation of the first-order
perturbation in Sec.~\ref{lowest}, we can
show that for a $N^{d-1}$-leg system on a $d$-D lattice, 
the lowest-magnon-gap slope also remains
finite at the $N\to \infty$ limit, 
like the case of ladders or tubes. 
For instance, the lowest gap of a weakly coupled 
integer-spin AF chains on $d$-D hypercubic lattice is evaluated as 
\begin{eqnarray}
\label{gap_hypercubic}
\Delta_{d}&=& \Delta - \frac{S^2Zv}{\Delta a_0}\sum_{a=2}^d
|J_\perp^{(a)}|,
\end{eqnarray}
where $J_\perp^{(a)}$ is the interchain exchange in $a$th direction (the
first direction is that of the chain).
Moreover, in the case of $J_\perp^{(a)}=J_\perp$, the slope of the gap is 
determined as $-d\Delta_d / d|J_\perp|=\frac{(d-1)S^2Zv}{\Delta a_0}$: 
in the spin-1, 3D case, the slope is $\simeq 15.3$, twice
$2\omega_\perp(0)/J_\perp$. 
Thus, our first-order perturbation theory suggests that a gapped phase
exists if all the interchain couplings are weak enough in any 
$d$-D spatially anisotropic integer-spin AF system. This prediction
is consistent with the general belief that gapped phases are 
robust against small perturbations.

\section{\label{decoration}Rung-coupling decorations}
\begin{figure}
\scalebox{0.5}{\includegraphics{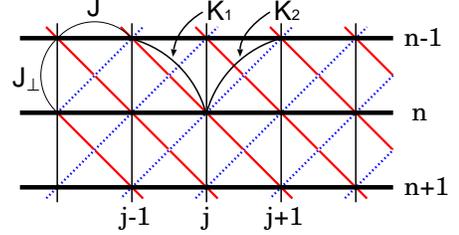}}
\caption{\label{fig4}Decorated rung couplings.}
\end{figure}

As we already mentioned, perturbation
theory, in principle, can deal with any sorts of interchain
interactions. Therefore, in this section, 
as a generalization of the standard coupling $\hat {\cal H}_\perp$, 
let us briefly investigate the following typical frustrated rung coupling:   
\begin{eqnarray}
\label{rung-dec}
\hat{\cal H}' &=& \hat{\cal H}_\perp 
+\hat{\cal H}_1 +\hat{\cal H}_2,\nonumber\\
\hat{\cal H}_1 &=& K_1\sum_{n,j}\vec S_{n,j}\cdot\vec S_{n+1,j+1},\nonumber\\
\hat{\cal H}_2 &=& K_2\sum_{n,j}\vec S_{n,j}\cdot\vec S_{n+1,j-1}.
\end{eqnarray}
This coupling is illustrated in Fig.~\ref{fig4}.
For example, an infinite-leg system with $K_1\neq 0$ and $K_2=0$ 
($K_1=K_2$) may be called a spatially anisotropic model on triangular lattice 
(spatially anisotropic $J_1$-$J_2$-like model). 
It is noteworthy that in contrast to the standard case of 
$\hat {\cal H}'=\hat {\cal H}_\perp$, 
when $\hat{\cal H}_{1({\rm or}\,2)}$ is present alone, odd-leg
{\em FM-rung} tubes exhibits geometrical frustration. 

As in the case of $\hat{\cal H}' = \hat{\cal H}_\perp$, 
the first-order correction of the ground-state energy, 
$\langle\Phi|\hat{\cal H}_{1,2}|\Phi\rangle$, are zero. Furthermore, the
effective Hamiltonians for $\hat{\cal H}_{1,2}$ in the
one-magnon space are easily obtained in the similar way from 
Eq.~(\ref{eff-H}) to Eq.~(\ref{eff-H-2}). The results are 
\begin{widetext}
\begin{eqnarray}
\label{eff-H-3}
\hat{\cal H}_{1,2}^{1\rm{st}} &\equiv& 
\hat {\cal P}_{\rm{one}}\hat{\cal H}_{1,2}\hat {\cal P}_{\rm{one}}
=\sum_\alpha\sum_{n=1}^{N'}\int\frac{vdp}{4\pi \omega(p)}
\left(-\frac{S^2 Zv}{2\omega(p)a_0}K_{1,2}\right)
\Big[|p,\alpha,n\rangle e^{\pm ipa_0}\langle p,\alpha,n+1|+
{\rm H.c.}\Big].
\end{eqnarray}
The exponential factors $\exp(\pm ipa_0)$ come from the fields 
$\vec m_n(x\pm a_0)$ in $\hat{\cal H}_{1,2}$. In ladder cases, 
if the rung coupling 
$\hat{\cal H}'$ contains only $\hat{\cal H}_1$ or $\hat{\cal H}_2$,
the factors could be replaced with unity by taking the field redefinition 
$\vec m_n(x\pm a_0)\to\vec{\tilde m}_n(x)$ at the intermediate stage in
the derivation of Eq.~(\ref{eff-H-3}). The replacement is physically
reasonable, because a ladder system with $\hat{\cal H}' = \hat{\cal H}_\perp$
is equivalent to one with 
$\hat{\cal H}' =\hat{\cal H}_{1\,({\rm or}\,2)}$ and 
$K_{1\,({\rm or}\,2)}=J_\perp$. However, one has to note that such a replacement 
procedure is not allowed when more than two kinds of rung couplings are
simultaneously present. Using Eq.~(\ref{eff-H-3}) and 
results of eigenvalue problems in Appendix~\ref{app1}, 
we obtain the following magnon bands 
of $N$-leg systems with a generalized rung coupling $\hat{\cal H}'$:  
\begin{eqnarray}
\label{general-band}
\tilde \omega_r^{1\rm st}(p) &=& \sqrt{\Delta^2+p^2v^2}
+\frac{S^2 Zva_0^{-1}}{\sqrt{\Delta^2+p^2v^2}}\times \left\{
\begin{array}{ccc}
|J_\perp-K_1 e^{-ipa_0}-K_2 e^{ipa_0}| \cos(k_r+\theta) & {\rm tubes}, \\
|J_\perp-K_1 e^{-ipa_0}-K_2 e^{ipa_0}| \cos k_r & {\rm ladders}, \\
\end{array} 
\right.
\end{eqnarray}
\end{widetext}
where the angle $\theta$ is defined by $J_\perp-K_1 e^{-ipa_0}-K_2 e^{ipa_0}
=|J_\perp-K_1 e^{-ipa_0}-K_2 e^{ipa_0}|e^{i\theta}$. 
In the dispersion relation, the phase of $e^{\pm ipa_0}$ may 
be negligible around $p=0$. 
While, when $p$ becomes too large (order of $\pi/a_0$), 
and thus the deviation between $e^{\pm ipa_0}$ and unity
becomes clear, the present NLSM picture for the $N$-leg system
is less reliable. 

In the case of $J_\perp=K_1+K_2$, the
magnon-band degeneracy around $p=0$ is not lifted by the
first-order correction of the rung coupling $\hat{\cal H}'$. 
This must be due to frustration among $\hat{\cal H}_\perp$,
$\hat{\cal H}_1$, and $\hat{\cal H}_2$. 
From this prediction, we can expect 
that the gapped-phase region in the infinite-leg system 
with a frustrated rung coupling are much larger than that in 
the standard system with $\hat{\cal H}'=\hat{\cal H}_\perp$ (see
Sec.~\ref{2D}). This is consistent with a recent work using
DMRG.~\cite{Mou}


\section{\label{Sum}Summary}
In this paper, utilizing only the information about 
integer-spin AF chains $\hat {\cal H}_0$ 
(exact results of the NLSM, and values of $\Delta$,
$v$, and $Z$), we have formulated a first-order perturbation 
expansion in the rung-coupling strength for the $N$-leg integer-spin
system~(\ref{N-leg}). All the magnon excitation energies have been 
quantitatively predicted as in Eqs.~(\ref{newband})-(\ref{trueband}) and
Fig.~\ref{fig_allgap}. Several features of the magnon excitation
structure have been elucidated [see comments (\rnum{1})-(\rnum{7}) in
Sec.~\ref{detail}]. 
The results of perturbation theory are 
supported by the QMC data and agree with our recent study 
based on NLSM and SPA in Ref.~\onlinecite{MS05}. 
We further applied the perturbation method to cases of $N\to\infty$ 
and systems with a generalized rung coupling in Secs.~\ref{2D} and 
\ref{decoration}, respectively. Particularly, in Sec.~\ref{2D}, it is
predicted that a Haldane state (gapped state with 1D nature) 
survives in spatially anisotropic
$d$-D ($d\geq 2$) integer-spin AF systems, if the exchange coupling of
one direction are sufficiently larger than all others [see
Eq.~(\ref{gap_hypercubic})]. This agrees 
with several previous studies based on different 
methods.~\cite{S-T,Matsu,Tasaki,Sene2,K-K}   
We expect that if fitting parameters
$\Delta$, $v$, and $Z$ are more accurately calculated, 
perturbation theory leads to almost exact predictions 
in the weak-rung-coupling regime.

The perturbation theory approach has some advantages over other methods. 
(\rnum{1}) It can lead to not
only the lowest magnon band, but also all other magnon ones. 
(\rnum{2}) One can predict several quantities with analytical
expressions if nonuniversal parameters $(\Delta,v,Z)$ are given. 
(\rnum{3}) Perturbation theory is expected to be applicable to all 
the systems with any leg number $N$. Therefore, as one found in
Secs.~\ref{Perturbation} and~\ref{2D}, 
the dimensional crossover behavior (how a 1D $N$-leg system approaches
the corresponding $d$-D one when $N$ becomes large) 
can be partially described. 
(\rnum{4}) It is possible to investigate 
any kind of interchain interactions.

We have thus demonstrated that the very simple perturbation theory can
give nontrivial results on weakly coupled integer-spin AF
chains. Within the first order, perturbation theory is reduced to an
eigenvalue problem in finite dimensions thanks to various selection
rules [see Eq.~(\ref{elements})]. The precise numerical data known for
the single chain then enables us to make quantitative predictions in the
weak-rung-coupling regime. 

Similar approach should also be useful for more general quasi-1D systems
consisting of weakly coupled {\it gapped} 1D systems. When the 1D system
is gapless, perturbation theory is not quite easy. Usually, we have
to rely on renormalization-group arguments, as in the case of the
well-studied two-leg ladder consisting of spin-$\frac{1}{2}$ AF chains.

\begin{acknowledgments}
The authors gratefully thank Munehisa Matsumoto for providing 
the QMC data in Table~\ref{tab2} and valuable discussions. 
M.S. thanks Takuji Nomura for useful comments.
This work is partly supported by a Grant-in-Aid for Scientific 
Research (B) (No. 17340100) from the Ministry of Education, 
Culture, Sports, Science and Technology of Japan. M.O. was supported 
in part by a Tokyo Institute of Technology 21st Century COE program 
``Nanometer-scale Quantum Physics,'' while he belonged to Department 
of Physics, Tokyo Institute of Technology until March 2006.

\end{acknowledgments}

\appendix

\section{Tridiagonal Matrices}
\label{app1}
Here, we briefly summarize solutions of eigenvalue problems of 
the following $N\times N$ Hermitian matrices,
\begin{subequations}
\label{ap1-1}
\begin{eqnarray}
{\cal A}&=& 
\left(
\begin{array}{cccccccc}
A_1        & A_2     &            &           &  A_2^*       \\
A_2^*      & A_1     &   \ddots   &           &          \\
           & \ddots  &   \ddots   &  \ddots   &          \\
           &         &   \ddots   &  \ddots   &  A_2     \\
A_2        &         &            & A_2^*     &  A_1     \\
\end{array}
\right),\label{ap1-1-1}\\
{\cal B}&=& 
\left(
\begin{array}{cccccccc}
B_1        & B_2     &            &           &         \\
B_2^*      & B_1     &   \ddots   &           &          \\
           & \ddots  &   \ddots   &  \ddots   &          \\
           &         &   \ddots   &  \ddots   &  B_2     \\
           &         &            &   B_2^*   &  B_1     \\
\end{array}
\right).\label{ap1-1-2}
\end{eqnarray}
\end{subequations}
The case in which $A_2$ and $B_2$ contain an imaginary part is not
discussed well in literature.

The eigenvalue problem for the matrix $\cal A$ ($N\geq 3$) 
is easily solved by the Fourier transformation method.  
The eigenvalues $\alpha_r$ and the corresponding eigenvectors 
$\vec A_r$ are given by 
\begin{subequations}
\label{ap1-2}
\begin{eqnarray}
\alpha_r &=& A_1+ 2|A_2|\cos
 (k_r+\theta),\,\,\,\,\,\,k_r=\frac{2r}{N}\pi,
\label{ap1-2-1}\\
\vec A_r&=&\sqrt{\frac{2}{N}}\,\,{}^T\Big(e^{ik_r},e^{2ik_r},
\dots,e^{Nik_r}\Big),\label{ap1-2-2}
\end{eqnarray}
\end{subequations}
where the angle $\theta$ is defined by $A_2=|A_2|e^{i\theta}$, 
$r=0,\dots, N-1$ (mod $N$), and $\vec A_r^2=1$. 
On the other hand, 
the matrix ${\cal B}$ can be diagonalized by assuming that 
the $n$th component of each eigenvector satisfies 
$b_n\propto e^{-in\phi}\sin(k n)$ where the angle $\phi$ is the argument
of $B_2$ ($B_2=|B_2|e^{i\phi}$) and $k$ is a constant.   
As a result, the eigenvalues $\beta_r$ and the eigenvectors $\vec B_r$ are 
\begin{subequations}
\label{ap1-3}
\begin{eqnarray}
\beta_r &=& B_1+ 2|B_2|\cos k_r , \,\,\,\,\,\, k_r=\frac{r}{N+1}\pi,\label{ap1-3-1}\\
\vec B_r&=&\sqrt{\frac{2}{N+1}}{}^T\Big(e^{-i\phi}\sin k_r,
e^{-i2\phi}\sin 2k_r,
\nonumber\\
&& \hspace{2cm}\dots,e^{-iN\phi}\sin Nk_r\Big),\label{ap1-3-2}
\end{eqnarray}
\end{subequations}
where $r=1,\dots,N$ and $\vec B_r^2=1$.

\section{Spin correlation functions in integer-spin AF chains}
\label{NLSMapp}
We demonstrate how the NLSM scheme derives the asymptotic 
form of spin correlation functions in the integer-spin AF chain 
$\hat{\cal H}_0$. As shown below, 
comparing the resulting form and that estimated by a
numerical method, the factor $Z$, etc. can be determined.

As we said in the main text, the low-energy physics of the chain 
$\hat{\cal H}_0$ is described by the NLSM~(\ref{actionNLSM}). 
From the formula~(\ref{spin}), the long-distance behavior of the 
equal-time two-point spin correlation is represented as 
\begin{eqnarray}
\label{corrNLSM}
\lim_{|j-k|\to\infty} \langle S_j^\alpha S_k^\alpha\rangle 
&\approx&
(-1)^{|j-k|}S^2\langle m^\alpha(x_j)m^\alpha(x_k)\rangle \nonumber\\
&&+\langle u^\alpha(x_j)u^\alpha(x_k)\rangle,
\end{eqnarray}
where $x_{j(k)}=j(k)\times a_0$. The term 
$\langle u^\alpha m^\alpha\rangle$ disappears due to 
the symmetry of $\vec m\to-\vec m$. Because $\vec m$ and $\vec u$ are
one-magnon and two-magnon fields respectively, the most relevant part 
at $|j-k|\to\infty$ is the first staggered term. Using the projection
operators in Eq.~(\ref{resolution}), we estimate it as follows: 
\begin{eqnarray}
\label{corrNLSM2}
\langle m^\alpha(x)m^\alpha(0)\rangle
&=& \Big\langle m^\alpha(x)
\Big[\sum_s\hat{\cal P}_s\Big] m^\alpha(0)\Big\rangle \nonumber\\
&=& \langle m^\alpha(x)
\hat{\cal P}_1 m^\alpha(0)\rangle +\dots\nonumber\\
&=& \frac{Zv}{4\pi}\int dp
 \frac{e^{ipx}}{\sqrt{\Delta^2+p^2v^2}}+\dots\nonumber\\
&=& \frac{Z}{2\pi}K_0(|x|/\xi)+\dots\nonumber\\
&\stackrel{x\to \infty}{\approx}& 
\frac{Z}{\sqrt{8\pi |x|/\xi}}\,\,e^{-|x|/\xi},
\end{eqnarray}
where $\xi=v/\Delta$ is the correlation length and $K_0(x)$ is the
modified Bessel function. If correct values of the spin-wave
velocity $v$ and the Haldane gap $\Delta$ are already known, 
the comparison between the spin correlation determined by 
Eqs.~(\ref{corrNLSM}) and (\ref{corrNLSM2}), and the one done by a 
numerical method can fix the parameter $Z$. Actually, the value of 
$Z$ in Table~\ref{tab1} is determined by such a method.~\cite{A-W,S-A}

In the similar way from Eq.~(\ref{corrNLSM}) to Eq.~(\ref{corrNLSM2}), 
the long-(imaginary)time behavior of the equal-position two-point spin 
correlation function is evaluated as
\begin{eqnarray}
\label{time-corr}
\langle S_j^\alpha(\tau) S_j^\alpha(0)\rangle 
&\approx& S^2\langle m^\alpha(x,\tau) m^\alpha(x,0)\rangle\nonumber\\
&\approx& \frac{S^2 Zv}{4\pi}\int dp\frac{e^{-\omega(p)\tau}}{\omega(p)}\nonumber\\
&=&  i\frac{S^2 Z}{4} H_0^{(1)} (i\Delta\tau)\nonumber\\
&\stackrel{\tau\to \infty}{\approx}& 
\frac{S^2Z}{\sqrt{8\pi\Delta\tau}}\,\,e^{-\Delta\tau},
\end{eqnarray}
where $\tau(=it)>0$ is the imaginary time, 
$S_j^\alpha(\tau)=e^{\hat{\cal H}_0\tau}S_j^\alpha e^{-\hat{\cal H}_0\tau}$, 
$m^\alpha(x,\tau)=e^{\hat{\cal H}_0\tau}m^\alpha(x) e^{-\hat{\cal H}_0\tau}$, 
and $H_0^{(1)}(x)$ is the Hankel 
function of the first kind. The form~(\ref{time-corr}) is an expected result from 
Eq.~(\ref{corrNLSM2}) (because the NLSM is a relativistic system).

\section{\label{Higher}Higher-order Terms}
We briefly consider higher than second-order corrections 
in the $N$-leg system~(\ref{N-leg}). In principle, following the
standard manner of perturbation theory, one would perform 
calculations up to any order. However, as one will see below,
there is a difficulty in extracting quantitative predictions from 
higher-order terms within the perturbation theory framework of the main
text.

Generally, a second-order perturbation term is represented as 
\begin{eqnarray}
\label{second}
\int \prod_{q=1}^Q\frac{vdp_q}{4\pi\omega(p_q)}
\langle A|\hat {\cal H}_\perp |Q\rangle  
\frac{1}{E_A-E_Q}\langle Q| \hat {\cal H}_\perp |A\rangle,
\end{eqnarray}
where $|A\rangle$ is an eigenstate of the unperturbed system 
$\hat {\cal H}_\parallel$ with an energy $E_A$ and a momentum $P_A$,
and similarly $|Q\rangle$ stands for a $Q$-magnon state with an energy 
$E_Q=\sum_q \omega(p_q)$ and a momentum $P_Q=\sum_q p_q$ ($p_q$
is the momentum of each magnon). The matrix element 
$\langle A|\hat {\cal H}_\perp |Q\rangle$ in the integrand
would contain
\begin{eqnarray}
\label{gene-element}
I(A,Q) J_\perp\int \frac{dx}{a_0} e^{i(P_Q-P_A)x},
\end{eqnarray}
where the factor $I(A,Q)$ is independent of $x$. 
Therefore, noticing that $\int dx e^{i(P_Q-P_A)x}=2\pi \delta(P_Q-P_A)$
and $\delta(P_Q-P_A)^2\sim L\times \delta(P_Q-P_A)$ where $L$ is the
chain length, we expect that a second-order term generally generates 
a quantity of $O(L)$. Actually, 
one can easily verify that in the second-order ground-state energy
correction, the term originating from 
the lowest-energy intermediate states 
with two magnons is $O(L)$.
Under the assumption that 
one-magnon states still take the lowest excitation 
when a finite rung coupling is added in decoupled chains 
$\hat{\cal H}_\parallel$, 
all $O(L)$ second-order perturbation terms for the ground state 
must completely cancel out those for one-magnon states 
at least in the thermodynamic limit ($L\to\infty$). 
Only $O(L^0)$ terms, if they are present, can contribute to the second-order
correction of the magnon-band splitting in the limit.

These arguments appear to suggest that almost all
the second-order perturbation terms are $O(L)$. However, 
the factor $I(A,Q)$ in Eq.~(\ref{gene-element}) can possess a
finite-size correction; for instance, the parameter 
$Z=|\langle 0|m^\alpha(0)|p,\alpha\rangle|^2$ might be expanded as 
$Z=Z(L)=Z_0+Z_1/L+Z_2/L^2+\dots$ ($Z$ in Table~\ref{tab1}
stands for $Z_0$). Furthermore, $v$ and $\Delta$ must also have a
finite-size effect. Such finite-size
corrections could generate an $O(L^0)$ term.
Unfortunately, as far as we know, 
finite-size correction terms have never been 
quantitatively estimated. Therefore, we cannot derive quantitative
predictions from the second-order perturbation expansion.

The similar difficulty also emerges 
in calculations of higher than third-order corrections.
Thus, we can conclude that it is generally hard to quantitatively 
calculate higher-order terms in the scheme in this
paper. Using the above expansion 
$Z(L)=Z_0+Z_1/L+\dots$, one might obtain 
some qualitative features of higher-order corrections 
in the magnon-band splitting.


\bibliography{apssamp}

\end{document}